\documentclass[a4paper,11pt]{article}
\pdfoutput=1 

\usepackage{jcappub} 

\usepackage[T1]{fontenc} 

\usepackage{subcaption,graphicx}

\usepackage{array}
\newcolumntype{F}[1]{>{ \centering \vspace{3mm} \arraybackslash}m{#1cm}<{ \vspace{3mm}\arraybackslash }}%

\title{\boldmath Natural Inflation with non minimal coupling to gravity in  $R^2$ gravity under the Palatini formalism}

\usepackage{float}

\author[a]{M. AlHallak}
\author[b]{N. Chamoun}
\author[a]{M. S. Eldaher}

\affiliation[a]{Physics Dept., Damascus University, Damascus, Syria}
\affiliation[b]{Physics Dept., HIAST, Damascus, Syria}
\emailAdd{phy.halak@hotmail.com}
\emailAdd{nidal.chamoun@hiast.edu.sy}
\emailAdd{m-saemaldahr@aiu.edu.sy}

\abstract{Natural Inflation with non-minimal coupling (NMC) to gravity, embodied by a Lagrangian term $\xi \phi^2 R $, is investigated in the context of an extended gravity of the form $R+ \alpha R^2$. The treatment is performed in the Palatini formalism. We discuss various limits of the model ``$\alpha \gg 1$'' and ``$\alpha \ll 1$'' in light of two scenarios of inflation: a ``Slow roll'' and a ``Constant roll'' scenario. By analyzing the observational consequences of the model, our results show a significant improvement regarding compatibility between the theoretical results of this model and the observational constraints from Planck 2018 and BICEP/Keck 2018, as exemplified by the tensor-to-scalar ratio and spectral index. Furthermore, a broader range for the parameter space of natural inflation is now compatible with the confidence contours of Planck \& BICEP/Keck results.
The joint effects of the contributions of both the NMC to gravity and the $\alpha R^2$ make a significant improvement: $\alpha R^2$ gravity influences scalar-tensor ratio values, whereas NMC to gravity has a more significant impact on the spectral index values. Contributions from both terms allow more previously excluded intervals to be included being compatible now with observational data. These conclusions about the roles of NMC to gravity and, particularly, the extended gravity remain mainly valid with a periodic NMC similar in form to the natural inflation potential. }
\makeatletter
\gdef\@fpheader{}
\makeatother
\begin{document}
\maketitle
\flushbottom
\section{Introduction}
\label{sec:intro}
The natural inflation (NI) scenario proposed by Freese, Frieman, and Olinto \cite{Freese:1990rb} is one of the most exciting models of inflationary cosmology due to some attractive features. First, the inflaton field of the model occurs naturally in particle physics because a spontaneously broken global symmetry results in the appearance of pseudo-Nambu-Goldstone bosons. A second property is its axion-like origin; thus, it possesses a shift symmetry with flat potential, preventing significant radiative corrections from being introduced into the potential.\par
The last property gives NI the ability to solve one of the theoretical challenges inherent to slow rolling inflation models, which are constrained by the fact that a relatively flat inflation potential slope is necessary to generate an adequate amount of inflation.
As a result of the flatness requirement of the potential slope, fine-tuning problems arise; in particular, quantum corrections in the absence of symmetry generally spoil the potential flatness, known as the $\eta$-problem \cite{Freese:1990rb}. NI models avoid this by utilizing an axionic field with a flat potential resulting naturally from a shift symmetry to drive inflation.
\par Despite having such intriguing features, NI with a cosine potential is disfavored at greater than $95 \% $ confidence level by current observational constraints from $Planck  2018$ on the scalar-tensor ratio $r$ and spectral index $n_s$ \cite{Planck:2018jri, Stein:2021uge}. However, a more recent analysis of BICEP/Keck XIII in 2018 (BK18) \cite{BK18} has put more stringent bounds on $r$, so we included in our observational data those of $Planck 2018$ (TT, EE, TE), BK18 and other experiments (lowE, lensing) separately or combined.\par
Moreover, this class of models can describe an inflaton field that is non-minimally coupled to gravity, with coupling $\xi \phi^2 R$. Non minimal coupling (NMC) to gravity is generally produced at one-loop order in the interacting theory for a scalar field, even if it is absent at the tree level \cite{Freedman:1974gs}. Actually, in general all terms of the form $(R^i \phi^j, R^{\mu\nu} \partial_\mu\phi \partial_\nu\phi, \ldots)$, which are covariant scalars and vanish in flat spacetime, are allowed in the action. However, omitting the derivative terms and taking a finite number of loop graphs enforce a polynomial form of the NMC term, and if one imposes CP symmetry on the action then the powers of the field $\phi$ therein are to be even. For simplicity, we included first only the quadratic monomial ($R \phi^2$) of dim-4, and studied its consequences in detail, whereas later, in the last section, we considered also an NMC of a periodic form respecting the shift symmetry of the NI potential.  An important role is played by NMC in the Higgs inflation model \cite{Spokoiny:1984bd,Bezrukov:2007ep,0809.2104} and Higgs stability problem (see \cite{Espinosa:2018mfn,Markkanen:2018pdo} for a brief review). On the other hand, current observational constraints on NMC are pretty loose, for example $|\xi|\leq 10^{15}$ from the LHC's result \cite{Atkins:2012yn}.

Unlike Palatini formalism, NI with NMC to gravity has been extensively studied in the Metric formalism. The work of \cite{Reyimuaji:2020goi} showed that NI with quadratic monomial NMC could in the metric formalism accommodate data. Moreover, the works of \cite{1806.05511,2002.07625}, adopting the metric formalism, showed that NI with periodic NMC could bring NI's predictions into a good agreement with Planck data, depending on values of the periodic NMC parameter $\lambda $ and the symmetry breaking scale $f$ specific to NI. The author of \cite{2107.03389} investigated in the metric formalism, within an extended gravity setup, NI with a periodic NMC, whose microscopic origin was proposed, and compared the predictions to Planck data. Also in metric formalism, in \cite{1907.00983} NI was combined with a specific quadratic tensor (the Weyl tensor squared) term and compared to Planck data, whereas in \cite{2202.00684} NI was combined with all possible quadratic-in-curvature terms in the action and compared to BK18 data.\par
On the other hand, the observations of type Ia supernovae (SNIa) \cite{SupernovaCosmologyProject:1998vns,SupernovaSearchTeam:1998fmf} show that the universe is currently at an accelerated expansion phase. This observed result contradicts our expectation from the behavior of ordinary matter, and it could not be explained based on general relativity (GR), putting aside the possibility of including a cosmological constant accounting for the accelerated expansion, but where the corresponding mechanism, albeit simpler than changing gravity or adding to the matter content, is not dynamic. One of the possible explanations for the current accelerated expansion of the universe is the modification of gravity law in such a way that it behaves as standard GR in strong gravitational regimes while acting as a repulsive force in the low-density cosmological scale.
As one of the most popular alternatives to general relativity, $F(R)$ gravity relies on modifications of the Einstein-Hilbert action by introducing nonlinear terms into the Ricci scalar $R$ \cite{Sotiriou:2008rp}. In this way, instead of using the Ricci scalar $R$, the gravitational Lagrangian becomes a well-defined function $F(R)$.\par
Obviously, this modification to gravity is not limited to the current epoch of Universe evolution, and it can be used to study different cosmological aspects \cite{Nojiri:2008nt,Elizalde:2010ep,Artymowski:2014gea,Contillo:2011fn}. The first full and internally self-consistent inflationary model developed within $F(R)$ gravity in the metric formalism was the pioneer paper of Starobinsky \cite{plb91} which remains viable by now. Cosmological inflation within  $F(R)$ gravity has since then been studied extensively \cite{Huang:2013hsb,Ketov:2010eg,Valtancoli:2018dnq,Sebastiani:2015kfa,Oikonomou:2021msx,Fomin:2020ndb,SBudhi:2019vln,Ketov:2013sfa,Hwang:2011kg}. Within the several formulations of $F(R)$ theories under the names of metric formalism, Palatini formalism and metric-affine formalism, the last one being the most general, the Starobinsky model of $F(R)$ in the Palatini formalism was studied in \cite{1608.03196} applied to the late expansion of the universe, and in \cite{1810.10418} applied to the early inflationary stage, whereas NI within Palatini formalism was studied in \cite{Antoniadis:2018yfq}.\par
This work aims to study the effects of both NMC to gravity and extended $\alpha R^2$ gravity on NI. A variety of scenarios are considered in this work. First, we consider a standard slow-roll inflation with a re-scaled scalar field and an effective potential. Secondly, we examine the limit $\alpha \gg 1 $ in which the model is reduced to a kinematically induced inflationary model (k-inflation) with a non-canonical scalar field.
Numerous studies were carried out to refine the k-inflationary scenario within the framework of the non-canonical scalar fields \cite{Garriga:1999vw, Armendariz-Picon:1999hyi,Helmer:2006tz,Panotopoulos:2007ky,Bose:2008ew,Devi:2011qm,Ohashi:2011na,Arroja:2011yj,Ohashi:2011ozd,Zhang:2011za,Ohashi:2013pca,Feng:2014pta,Peng:2016yvb,Sebastiani:2017cey,Do:2020hjf,Pareek:2021lxz}. Inflationary models within string theory exhibit unusual scalar field dynamics involving non-minimal kinetic terms \cite{Ringeval:2009jd,Langlois:2009ej}, whereas \cite{Taveras:2008yf} considered loop-quantum gravity-inspired modifications on GR. The Holst action in this latter case is generalized by making the Barbero-Immirzi (BI) parameter a scalar field, whose value could be dynamically determined. The modified theory leads to a non-zero torsion tensor that corrects the field equations through quadratic first-derivatives of the BI field. In our work, the modified gravity represented by NMC to gravity, in addition to an extended $\alpha R^2$ gravity, produces the non-canonical kinetic term for the scalar field \cite{Oikonomou:2021edm,Enckell:2018hmo}.

In a final point, we investigate the limit $\alpha \ll 1$ within the so-called constant-roll inflation, specified by a parameter $\beta$ to remain constant during the inflation (see \cite{Motohashi:2014ppa,1702.05847,Odintsov:2017hbk,Odintsov:2017qpp,Oikonomou:2017xik,Oikonomou:2017bjx,Nojiri:2017qvx,Awad:2017ign,
Anguelova:2017djf,Karam:2017rpw,Mohammadi:2018wfk,Motohashi:2019tyj,Odintsov:2019ahz,Mohammadi:2018zkf,Mohammadi:2019qeu,
Antoniadis:2020dfq} where the confrontation of the model, within arbitrary $\beta$, in GR with recent observational data has been first made). On the other hand, the work of \cite{Kinney:2005vj} treating the case ($\beta = 3$) is usually cited in connection with ``ultra-slow-roll" inflation. Different types of inflation were re-examined in the context of a constant roll scenario. Tachyon inflation is studied with a constant rate of rolling \cite{Gao:2018tdb,Mohammadi:2018oku}, where the authors derived to first order the analytical expressions for the scalar and tensor power spectra, the scalar and tensor spectral tilts, and the tensor to scalar ratio by using the method of Bessel function. A constant roll scenario within modified gravity was investigated in \cite{Eurphys} where all constant-roll inflationary models in $F(R)$ gravity in metric formalism were found in analytical form.  \par
In \cite{AlHallak:2021hwb}, we studied the effect of both NMC to gravity and an extended $\alpha R^2$ gravity applied to an inflationry model inspired by ``variation of constants'' leading to an exponential potential of a certain form, and found that the joint effects of both factors allowed an accommodation with data. On the other side, we apply the same techniques in this work on the NI and see that both effects allow a sort of ``focusing'" of the predictions, so that increasing ``$\alpha$'' pushes the predictions downward towards smaller $r$ whereas increasing the NMC parameter ``$\xi$" pushes the predictions rightward towards larger $n_s$, and that both factors play in a disjoint way of each other, in such a way that one can meet the observational constraints, even if they were tight.

At last, we investigate the case of periodic NMC to gravity, in the slow roll regime of the inflationary paradigm, where the fact that $\phi$ is a pseudo Nambu-Goldstone boson with a residual shift symmetry respected by the potential can force the NMC term to have the same shift symmetry \cite{2107.03389}. Here, we find that the conclusions of the NMC quadratic monomial ($\xi \phi^2 R$) are in line with those for a periodic NMC.

The paper is organized as follows.  In section 2, we introduce the model in the Palatini formalism and we apply the conformal transformation in which the model is converted to a scalar field model $\mathcal{F}(X)$, where $X$ is the canonical kinetic term of the field $\phi$, with an effective potential. We study in section 3 the slow roll regime, where the analysis strategy is defined, followed by a comparison with observations. In section 4, we treat the case of large $\alpha $ where the model reduces to K-inflation. Observables of $n_s$ and $r$ are computed and compared with the experimental data of Planck 2018 \& BK18. Section 5 is devoted  to the study of small values of $\alpha$, and the model at this limit is treated within the constant roll scenario. In section 6, we treat the case of a periodic NMC and end up with a summary and conclusion in section 7.

\section{The Model}
\label{sec:Model}
We consider the axions to play the role of a scalar field $\phi(x)$ that is non-minimally coupled to gravity via its couplings to the Ricci scalar  $R(g, \Gamma)$. We perform the study within an extended gravity represented by $R+\alpha R^2 $. A most general action would be
\begin{equation}
\label{eq:GAction}
S=\int \sqrt{-g}\bigg\{\frac{1}{2} R + \frac{1}{2} \alpha R^2 + \frac{1}{2} F(\phi) R - \frac{1}{2} g^{\mu\nu}\partial_\mu \phi \partial_\nu \phi - V(\phi)   \bigg\} .
\end{equation}
Here, $g$ is the determinant of the metric $g_{\mu\nu}$ and $R = g^{\mu\nu}R_{\mu\nu}(\Gamma,\partial \Gamma)$ is the Ricci scalar, where the Ricci tensor $R_{\mu\nu}$ is built out of the connection $\Gamma$ which would be independent of the metric in the Palatini formalism.
$V(\phi)$ is the potential driving the NI which is invariant under a shift symmetry $\phi \rightarrow \phi+ 2 \pi f$ \cite{Freese:1990rb}. In the original treatment of NI, the potential is
\begin{equation}
\label{eq:potential}
V(\phi)=M^4 \bigg(1+\cos(\frac{\phi}{f})\bigg) .
\end{equation}
where $M$ is a scale of an effective field theory generating this potential, $f$ is a symmetry breaking scale \cite{Peccei:1977hh}.

As mentioned in the introduction, the form of the function $F(\phi)$ is taken to be  \begin{equation} \label{quadratic-monomial}F(\phi)= -\xi \phi^2\end{equation} generated by quantum corrections. Actually, following \cite{conformal}, one can motivate such a form considering that for a massless scalar field in a curved spacetime, the corresponding Klein-Gordon equation needs, in order to be conformally symmetric, to include such a term for a specific value $\xi = \frac{1}{6}$. Taking into account the ``not yet computed" loop quantum effects, depending on the ``unknown" microscopic theory, pushes us to take $\xi$ as a free parameter. Moreover, in \cite{Reyimuaji:2020goi}, simplifying arguments, and CP conservation assumption, were presented to justify the form of Eq. (\ref{quadratic-monomial}). However, in the last section, we shall also investigate, albeit more briefly, the model with a periodic NMC to gravity of the same form as the periodic potential (Eq. \ref{eq:potential}), i.e. ($F(\phi) = \lambda \left(1+\cos (\frac{\phi}{f})\right)$).

It has been shown that either one of the NMC to gravity term or the extended $F(R)$ gravity term would lead to some interesting results concerning inflation. The effect of each one of these corrections has been studied recently\cite{SupernovaSearchTeam:1998fmf,Antoniadis:2018yfq}. In this work, we shall study the joint effect of both terms.

We can write the above effective classical action in terms of an auxiliary scalar $\chi$ as follows:
\begin{equation}
\label{eq:S_action}
S=\int d^4x\sqrt{-g}\bigg\{\frac{1}{2}(\chi-\xi \phi^2)R-\frac{(\chi-1)^2}{8\alpha}-\frac{1}{2} g^{\mu\nu}\partial_\mu \phi \partial_\nu \phi - V(\phi)\bigg\}
\end{equation}
Unlike working within the metric formalism \cite{Sotiriou:2008rp,Kubota:2011re}, the field $\chi$ has no kinetic term. As a result, its equation of motion reduces to a constraint on an action that describes a single scalar field $\phi$ non-minimally coupled to gravity.

It is possible to proceed with this form of the Lagrangian in what is called the Jordan frame. However, suppose we switch to the Einstein frame and use variables with minimal coupling between the scalar field and gravity. In that case, we will be able to use the familiar equations of GR, the inflationary solutions, and the standard slow-roll analysis. Making the conformal transformation (see for e.g. \cite{Kubota:2011re})
\begin{equation}
\label{eq:conf_trans}
\tilde{g}_{\mu\nu}=\Omega^2(\xi,\phi,\chi)g_{\mu\nu}
\end{equation}
where $\Omega^2=(\xi \phi^2 + \chi)$, then
the action, after dropping the tilde on $g_{\mu\nu}$ thereafter, becomes
\begin{equation}
\label{eq:sActionWchi}
S=\int d^4x \sqrt{-g}\bigg\{\frac{1}{2}R-\frac{1}{2} \frac{1}{\chi - \xi \phi^2}g^{\mu\nu}\partial_\mu \phi \partial_\nu \phi - \tilde{V}\bigg\}
\end{equation}
where the potential becomes:
\begin{equation}
\label{Vtilde}
\tilde{V}=\frac{8 \alpha V(\phi)+(\chi-1)^2}{8\alpha\Omega^4}
\end{equation}
As we mentioned before, the auxiliary field $\chi$ acts as a constraint on the model. Varying \eqref{eq:sActionWchi} with respect to $\chi$, we can find the relation between the scalar field $\phi$ and the auxiliary field $\chi$ as:
\begin{equation}
\label{chi_phi}
\chi=\frac{1-\xi \phi^2 +8\alpha V + 2 \alpha \xi \phi^2  \partial^\mu \phi \partial_\mu \phi}{1-\xi \phi^2 - 2 \alpha \partial^\mu \phi \partial_\mu \phi}
\end{equation}
By substituting \eqref{chi_phi}  into  \eqref{Vtilde} and \eqref{eq:sActionWchi}  we can write the action in terms of the scalar field $\phi$
\begin{equation}
\label{Action_w_XX}
S=\int d^4x \sqrt{-g}\bigg\{\frac{1}{2}R+ \frac{X}{(1-\xi \phi^2)(1+8\alpha \hat{V})}+\frac{2\alpha X^2}{(1-\xi \phi^2)^2(1+8\alpha \hat{V} )}-W(\phi)\bigg\}
\end{equation}
where we have defined
\begin{subequations}\label{eq:y}
\begin{align}
\label{eqOfX}
X \equiv - \frac{1}{2} g^{\mu\nu}\partial_\mu \phi \partial_\nu \phi
\\
\label{Vhat}
\hat{V}\equiv\frac{{V}}{\left(1-\xi \phi^2\right)^2}
\end{align}
\begin{equation}
\label{Veff}
W\equiv\frac{\hat{V}}{1+8\alpha \hat{V}}
\end{equation}
\end{subequations}

The action \eqref{Action_w_XX} shows that the contribution of the $R^2$ term turned the model into non-canonical scalar field model with kinetic term of the form $\mathcal{F}(X,\phi)=g_1(\phi) X + g_2(\phi) X^2 $ with an effective potential $W$. In the following sections, we will investigate the different limits of the model in different inflation scenarios.

\section{The Slow Roll Inflation Case : $2 \alpha X \ll 1-\xi \phi^2$}
\label{slow_roll_Inflation}
\subsection{Formulation}

Consider the slow-roll inflation scenario with `small' kinetic energy dominated by the potential energy, and so we can ignore the terms of higher-order kinetic energy. By defining a new scalar field $\varphi$ as
\begin{equation}
\label{phivar}
Z=\bigg(\frac{d \phi}{d \varphi}\bigg)^2 = (1+8 \alpha \hat{V})(1-\xi \phi^2)
\end{equation}
we see that the model is recast in a standard slow-roll inflation form with an effective potential $W(\phi(\varphi))$ \eqref{Veff}.

\begin{figure}
  \begin{subfigure}{0.45\textwidth}
    \includegraphics[scale=.55]{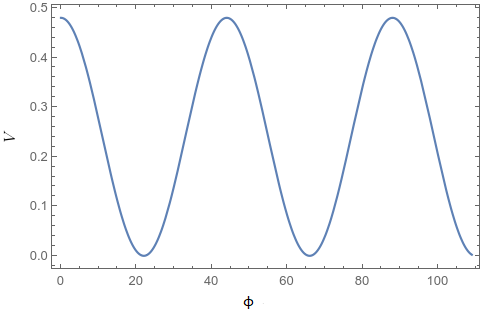}
    \caption{} \label{V_phi}
  \end{subfigure}%
  \hspace*{\fill}   
  \begin{subfigure}{0.58\textwidth}
    \includegraphics[scale=.5]{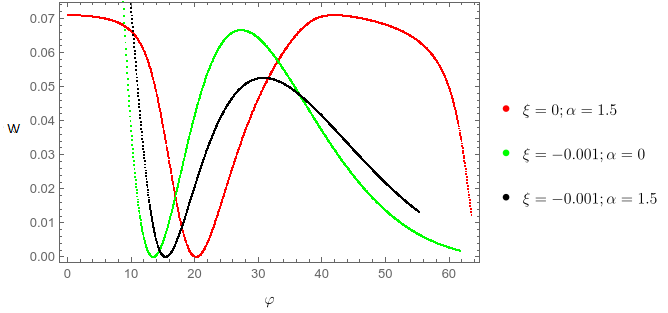}
    \caption{} \label{U_varphi}
  \end{subfigure}
\caption{(left panel) Ratio of the potential to fourth power of its scale, $V/M^4$, as a
function of original field $\phi$ in the Palatini formulation of gravity. (right panel) Ratio of the potential to fourth power of its scale. The changes with different values of the coupling constant $\xi$ and the parameter $\alpha$ are indicated by different color lines. In both panels, we took $f=16$, in line with values adopted in \cite{Antoniadis:2018yfq,2004.05238}.}
\label{V_and_U_wrt_varphi}
\end{figure}

To have a better intuitive understanding of the properties of the potential $W$ \eqref{Veff} in the Einstein frame, we plot it as a function of the canonically normalized field $\varphi$ in Fig. (\ref{V_and_U_wrt_varphi}),  with $f=16 M_p$, $M \sim M_{GUT}$.  The figure shows three different cases: the green line corresponds to NMC to gravity case with $\xi = -0.001 $ and $\alpha =0$, where the effect appears as a flattening and damping factor on the periodic potential $V$; the red line represents the minimal coupling to an extended gravity with  $\alpha=1.5$ and $\xi=0$, and the main impact is reflected in producing a flattened Einstein frame potential  $W \sim \frac{1}{(8 \alpha)}$ for non-negligible values of the periodic potential $V$; finally, the black line indicates both NMC to gravity and extended gravity with $\alpha=1.5$ and $\xi=-0.001$.

\begin{figure}[tbp]
\centering
\includegraphics[width=13cm, height=8cm]{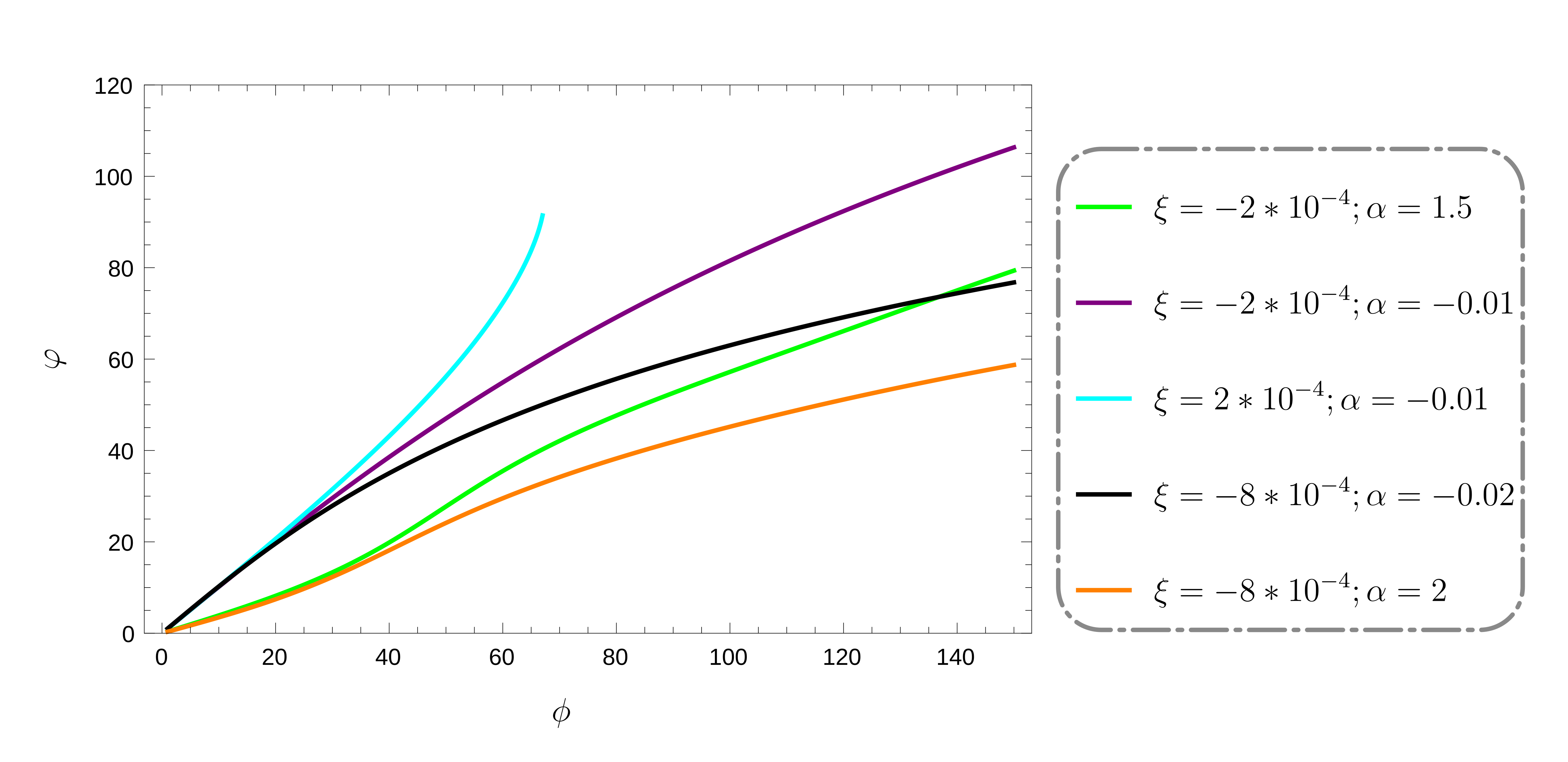}
\caption{\label{fig:varphi_with_phi} Plot of field $\varphi$ as a function of $\phi$ with $f=16$ and $M=0.7$ (choosing a positive slope of $\frac{d\phi}{d\varphi}=\sqrt{Z}$). Different color lines indicate different values of the coupling $\xi$ and $\alpha$}.
\end{figure}

One can not integrate Eq. (\ref{phivar}) to obtain an analytic expression of $\varphi(\phi)$, let alone invert it to get $\phi(\varphi)$. However, the relationship between $\varphi$ and $\phi$ is found numerically and is represented in Fig.  (\ref{fig:varphi_with_phi}), in which we opted to take the positive square root of Eq. (\ref{phivar}), and thus $\varphi$ is a growing function of $\phi$ with positive slope.

The potential \eqref{Veff}, obtained numerically, reveals two regimes: ‘large field inflation' in which $\varphi$ (or $\phi$) rolls from right leftwards for $\phi > \frac{1}{\sqrt{|\xi|}}$, and ‘small field inflation' in which $\varphi$ (or $\phi$) rolls from left rightwards for $\phi < \frac{1}{\sqrt{|\xi|}}$. In our study we shall be concerned with the small field paradigm, and so the field rolls from left to right up to where the potential $W$ reaches a minimum ($\frac{dW}{d\varphi}=\sqrt{Z} \frac{dW}{d\phi}$), which corresponds to $\phi = \frac{1}{\sqrt{|\xi|}}$ (for negative $\xi$) making $Z$ of Eq. (\ref{phivar}) to vanish. The inflationary regime ends by oscillations near this minimum.

The resultant action in Einstein frame has gravitational part of the from of Hilbert-Einstein action in addition to canonical scalar field $\varphi$ with an effective potential $W(\phi(\varphi))$, so we can apply the standard slow roll analysis characterized by the slow-roll parameters
\begin{equation}
\label{Non_physical_SRP}
\varepsilon=\frac{1}{2}\bigg( \frac{W_\varphi}{W}\bigg)^2 ,\qquad  \eta=\frac{W_{\varphi\varphi}}{W}
\end{equation}
The same quantities can be formally defined for the field $\phi$ but are not of physical meaning. However, it is easier to compute  the physical slow-roll parameters using the differentials with respect to $\phi$ via
\begin{equation}
\label{physical_eps}
\varepsilon=\frac{Z}{2(1+8\alpha \hat{V})}\bigg( \frac{\hat{V}_\phi}{\hat{V}}\bigg)^2
\end{equation}
\begin{equation}
\label{physical_eps}
\eta=\frac{Z}{(1+8\alpha \hat{V})}\bigg( \frac{\hat{V}_{\phi\phi}}{\hat{V}}\bigg)+\frac{Z_\phi}{2(1+8\alpha \hat{V})}\bigg(\frac{\hat{V}_\phi}{\hat{V}}\bigg)
-\frac{16\alpha Z \hat{V}}{(1+8\alpha \hat{V})^2}\bigg(\frac{\hat{V}_\phi}{\hat{V}}\bigg)^2
\end{equation}

\begin{figure}[tbp]
\centering
\includegraphics[width=13cm, height=10cm]{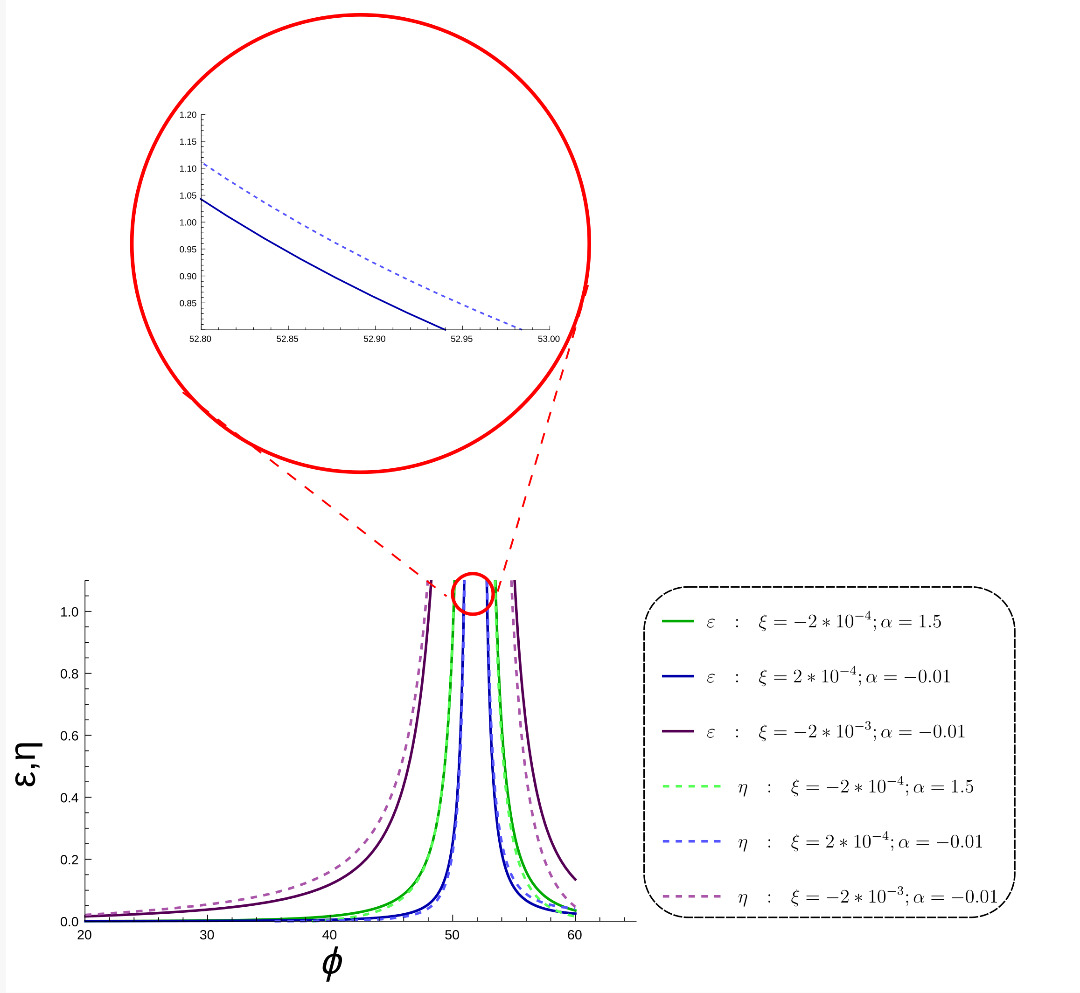}
\caption{\label{eta_vs_epsilon} The slow-roll parameters $\varepsilon$ and $\eta$ for different values of pamarmters $\xi$ and $\alpha$ }
\end{figure}

For comparative purposes, we draw in Fig. (\ref{eta_vs_epsilon}) the plots of these slow-roll parameters. We see that both parameters follow a similar pattern; however, we examined which parameter reached unity first, despite the lack of significant differences. Although we have analytical expressions for the slow-roll parameters, they are not easy to solve in order to determine their root value $\phi_\ast$ at which the observables $ n_s $ and $ r $ should be evaluated. So instead, we proceed numerically by finding $\phi_{end}$ which is determined by $\eta,\epsilon = 1$ according to the one that satisfies the condition first, and then going a fixed value of $N=\int_{t_i}^{t_e} Hdt = -8\pi G \int_{\varphi_i}^{\varphi_e} \frac{W}{W_\varphi} d\varphi =  -8\pi G \int_{\phi_i}^{\phi_e} \frac{W}{Z W_\phi} d\phi \in [50,80]$ e-folds back to obtain $\phi_\ast$ while making sure that the slow-roll parameters remain small in this range of $\phi$.

Accordingly, the spectral index of the primordial curvature perturbations is equal to
\begin{equation}
\label{spectralIndex_I}
n_s= 1 - 6 \varepsilon + 2 \eta
\end{equation}
In addition, the analytic form of the tensor-to-scalar ratio $r$ is
\begin{equation}
\label{scalar_tensor_I}
r=16 \varepsilon
\end{equation}

\subsection{Comparative analysis with observation:}

\begin{figure}
  \begin{subfigure}{0.49\textwidth}
    \includegraphics[scale=.45]{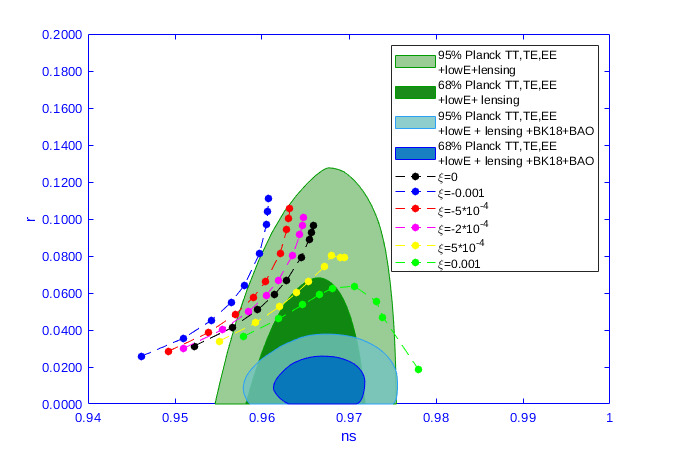}
    \caption{} \label{comparision_with_NMC_within_palatini:a}
  \end{subfigure}%
  \hspace*{\fill}   
  \begin{subfigure}{0.49\textwidth}
    \includegraphics[scale=.45]{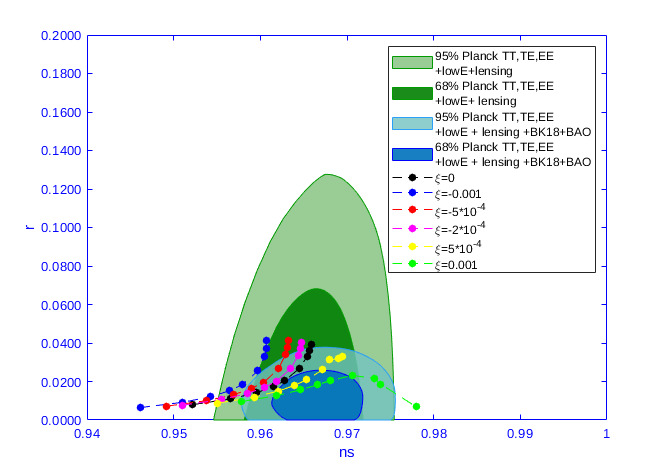}
    \caption{} \label{comparision_with_NMC_within_palatini:b}
  \end{subfigure}

\caption{The left figure (4.a) shows the  predictions of the non-minimally coupled natural inflation (NMCNI) with $\alpha = 0$.  All dotted curved lines correspond to N=60. The right figure (4.b) represents the results obtained for extended NMCNI with $\alpha =0.1$ } \label{comparision_with_NMC_within_palatini}
\end{figure}

For comparative purposes we reproduce the results of ref \cite{Reyimuaji:2020goi} in Fig. (\ref{comparision_with_NMC_within_palatini:a}) corresponding to the predictions of the NI model with NMC to gravity for the scalar spectral index $n_s$ and the tensor-to-scalar ratio $r$. All lines correspond to the $N=60$ e-folds with $\alpha=0$ (no extended gravity). For every curved line, the parameter $f$ domain is $f \in [\,1.5-10]\,$. The value of $\xi$ is fixed for each dotted curve and we took these $\xi$ values to range from $+0.001$ (green) to $-0.001$ (blue). As $\xi$ increases, the curved lines tend to shift toward more significant values of $n_s$. At relatively low values of $f$ and $\xi <0$, the lines penetrate the $95\%$ confidence level (CL) region of the combined Planck TT-TE-EE+LowE+lensing. However, the lines do not intersect with the corresponding new observational constraints upon adding the BK18 results. The larger the values of $|\xi|$ are, the more the corresponding lines shy away from the $68\%$ CL region of Planck TT-TE-EE+LowE+lensing.
On the other hand, for $\xi >0$ and  as  $\xi$ increases, the spectral index is pushed to the right approaching, but still not reaching, the $95\%$ CL regions of Planck $\&$ BK18 combined. However the $68\%$ CL region is attained when excluding BK18. In brief, with no extended gravity, there is hardly a good agreement with data when excluding BK18, in that one touches the $68\%$ CL only for $\xi>0$, and the disagreement gets worse if the BK18 data are included, in that even the $95\%$ CL region can not be reached.

The effect of adding the term $\alpha R^2$ can be seen in Fig. (\ref{comparision_with_NMC_within_palatini:b}). The same domains for the parameters $f$ and $\xi$ are scanned but now for non-zero $\alpha=0.1$. The primary influence of the $\alpha$ exhibited on the tensor-to-scalar ratio  is that higher values of $\alpha$ are associated with lower tensor-to-scalar ratio values. Consequently, the $95\%$ portions of the lines ``in the case of $\alpha=0$" tend now to be included in both the  $68\%$ and $95\%$ CL regions. Also, a broader range of previously excluded parameters $f$ and $\xi$ can now be included, at least for the $95\%$ region. Note, also, that the imposition of the new constraints originating from BK18 can be met here with $\alpha \neq 0$ in the case of $95\%$ CL for both signs of $\xi$, and in the case of $68\%$ CL for $\xi>0$ only.

\begin{figure}[tbp]
\centering
\includegraphics[scale=.8]{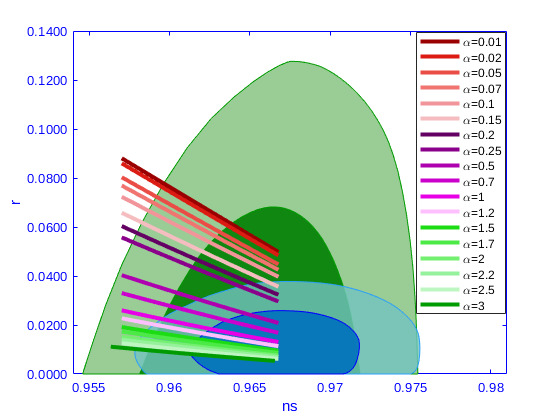}
\caption{\label{fig:xi_zero_alpha_positive} A plot of $r$ and $n_s$ for minimally coupled  NI with extended $\alpha R^2$ gravity. Each line corresponds to a fixed value for $\alpha$ and to scanned vlaues of $N$. Number of e-foldings $N$ is scanned form $50$ at left boundaries of the lines to $70$ at right boundaries}.
\end{figure}

\begin{figure}
  \begin{subfigure}{0.49\textwidth}
    \includegraphics[scale=.6]{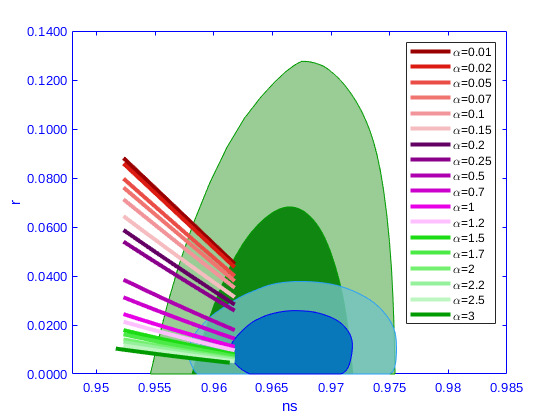}
    \caption{} \label{xi_negative_alpha_positive:a}
  \end{subfigure}%
  \hspace*{\fill}   
  \begin{subfigure}{0.49\textwidth}
    \includegraphics[scale=.6]{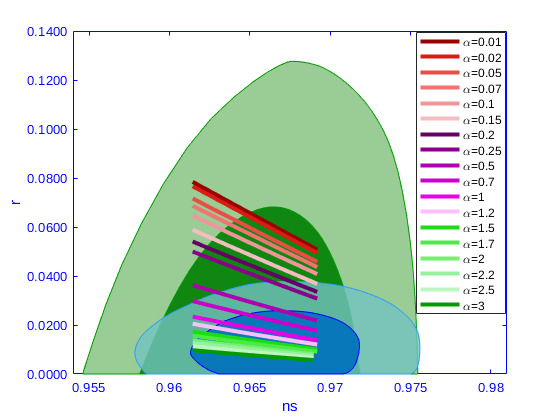}
    \caption{} \label{xi_negative_alpha_positive:b}
  \end{subfigure}
  \caption{Contour plots of $n_s$ and $r$ in the case of NMC scalar field within $\alpha R^2$ gravity. Left panel corresponds to $\xi=-1\times 10^{-3}$ while right panel corresponds to $\xi=+5\times 10^{-3}$, whereas other parameters are identical to those of Fig. \ref{fig:xi_zero_alpha_positive}. } \label{fig:xi_post_negat_alpha_positive}
\end{figure}

As a way of better understanding the effect of NMC to gravity along with extended $\alpha R^2$ gravity, we reproduced the results of ref \cite{Antoniadis:2018yfq}, corresponding to minimal coupling to gravity, in Fig. (\ref{fig:xi_zero_alpha_positive}). When $M$ and $f$ are given, the variation of $\alpha$ influences only the tensor-to-scalar ratio $r$. In agreement with previous findings, increasing $\alpha$ leads to smaller values of $r$ \cite{Enckell:2018hmo} which helps in accommodating data, in that without the $R^2$ term ($\alpha = 0$), one does not achieve a good agreement with data especially because of the stringest upper bounds on $r$ put by (BK18), whereas allowing for $R^2$ term ($\alpha \neq 0$) allows for redusing $r$ without affecting much the $n_s$ values, leading thus to accommodation of data. In Figs. (\ref{xi_negative_alpha_positive:a}) and(\ref{xi_negative_alpha_positive:b}), we examined the effect of NMC term on the $\alpha R^2$ extended gravity results. With increasing positive values of parameter $\xi$, the lines displace to larger values of $n_s$ with slight impact on $r$. In consequence, a better agreement with observations is achieved. Likewise, decreasing negative values of $\xi$ have the opposite impact on $n_s$. Eqs. (\ref{schema1}, \ref{schema2}) summarize the situation:
\begin{eqnarray}
\xi \nearrow & \Rightarrow & n_s \nearrow \label{schema1} \\
\alpha \nearrow & \Rightarrow & r \searrow \label{schema2}
\end{eqnarray}

Similarly, the joint existence of the terms of $\alpha R^2$ and $\frac{1}{2}\xi R \phi^2$, having distinct and separate impacts, lead to a tremendous implication for the theoretical results of NI. As discussed, the influence of the first term is restricted mainly to the $r$ values, while the second term concerns essentially  the  $n_s$ values. Since the parameters $\xi$ and $\alpha$ are pretty loose, this gives the model the ability to fit the higher accurate future observations of $r$ and $n_s$, since having both terms together serves as a ``focusing" tool affecting separately the $r$ and $n_s$ values.

Table \ref{table_1} shows the effect of both terms leading to include some previously excluded points within observations.
\begin{table}[ht]
\centering
\caption{ Benchmark points with/without the NMC to gravity and $R^2$ extension}
\begin{tabular}{ |F{.7}|F{.7}|F{.9}|F{1.8}|F{1.3}|F{1.5}|F{6.35}| }
 \hline
 $f$ & $N$ & $\alpha$ & $\xi$ & $n_s$ & $r$ & Statues\\
 \hline
 $5.0$ & $60$ & $0.00$& $0$ & $0.9522$ & $0.0313$ & Excluded \\
 $5.0$ & $60$ & $0.00$ & $1 \times 10^{-3}$ & $0.9579$ & $0.0365$ & $2\sigma$ TT+TE+EE+LowE+lensing \\
 $7.0$ & $50$ & $0.01$&$0$ & $0.9570$ & $0.0881$ & Excluded \\
 $7.0$ & $50$ & $0.01$ &$5\times 10^{-4}$ & $0.9614$ & $0.0784$ &$ 2\sigma$ TT+TE+EE+LowE+lensing\\
 $10$ & $60$ &$0.00$ &$-5\times 10^{-4}$ & $0.9632$ & $0.1056$ &  $2 \sigma$ TT+TE+EE+LowE+lensing \\
 $10$ & $60$ &$0.10$&$-5\times 10^{-4}$ & $0.9632$ & $0.0595$ &  $1 \sigma$ TT+TE+EE+LowE+lensing \\
 \hline
\end{tabular}
\label{table_1}
\end{table}%

\section{K inflation case:$\alpha \gg 1$}
\label{k-inflation}
\subsection{Analysis}

In this section, we will treat the model as a K-inflation scenario by taking into consideration the contribution of the square of the kinetic energy $X$, and assuming it dominates over the other linear term in $X$. Thus, we shall take the limit,
\begin{equation}
\label{limit_case_II}
1 \ll \frac{2 \alpha X}{(1-\xi \phi^2)}
\end{equation}
and, as a result, the Lagrangian of Eq. (\ref{Action_w_XX}) takes the form:
\begin{equation}
\label{lagrangian_caseII_phi}
L=\frac{R}{2} + P(\phi) X^2 + W(\phi)
\end{equation}
where we have $P(\phi)=\frac{2\alpha}{(1-\xi \phi^2)^2(1+8 \alpha \hat{V})}$. By defining a new scalar field $\varphi$ via
\begin{equation}
\label{psi_II}
\frac{\partial \varphi}{\partial \phi}= P^{1/4}(\phi),
\end{equation}
the model takes the form.
\begin{equation}
\label{Action_canonical_psi}
S=\int d^4x \sqrt{-g}\bigg\{\frac{1}{2}R+ \Big(\frac{1}{2}\partial_\mu \varphi \partial^\mu \varphi\Big)^2-W\bigg\}
\end{equation}
Friedmann equations for this case are given as \cite{1204.6214}:
\begin{equation}
\label{Fried_1_k_infl}
3 H^2=\frac{3}{4}\Dot{\varphi}^4 + W(\varphi)
\end{equation}
whereas the equation of motion of the scalar field $\varphi$ is,
\begin{equation}
\label{EOM_varphi_K_infl}
3 \ddot{\varphi} + 3 H \dot{\varphi} + W_\varphi \dot{\varphi}^{-2}=0
\end{equation}
The spectral index $n_s$ and the tensor-to-scalar ratio $r$ are given \cite{1204.6214} now by
\begin{equation}
\label{r_caseII}
r=\frac{16}{3} \varepsilon
\end{equation}
\begin{equation}
\label{ns_caseII}
n_s-1=\frac{1}{3} (4\eta - 16 \varepsilon)
\end{equation}
where
\begin{equation}
\label{varespilon_caseII}
\varepsilon=\frac{1}{2} 3^{\frac{1}{3}}\frac{(W_\varphi)^{\frac{4}{3}}}{W^{\frac{5}{3}}}
\end{equation}
\begin{equation}
\label{eta_caseII}
\eta= 3^{\frac{1}{3}}\frac{(W_{\varphi\varphi})}{(W.W_\varphi)^{\frac{2}{3}}}
\end{equation}
As we mentioned before, all derivatives could be taken with respect to the scalar field $\phi$ provided we do include the factor
\begin{equation}
\label{canonical_field_case_II}
\left(\frac{\partial \phi}{\partial \varphi}\right)^2= Z \equiv \Bigg(\frac{(1-\xi \phi^2)^2+8 \alpha V }{2\alpha}\Bigg)^{\frac{1}{4}}
\end{equation}
The analytical expression of ($n_s, r$) are complicated. However in the case of minimal coupling to gravity, they are relatively simple:
\begin{equation}
\label{analytical_exp_ns}
n_s-1=\frac{4 \left(\frac{2}{3}\right)^{2/3} \cos ^2\left(\frac{\phi}{2 f}\right) \left(3 \alpha M^4 \cos \left(\frac{2 \phi}{f}\right)+(1-8 \alpha M^4) \cos \left(\frac{\phi}{f}\right)-11 \alpha M^4-2\right)}{\sqrt[3]{\alpha} f^{4/3} \sqrt[3]{M^4} \sin ^{\frac{2}{3}}\left(\frac{\phi}{f}\right) \left(\cos \left(\frac{\phi}{f}\right)+1\right)^{5/3} \left(8 \alpha M^4 \cos \left(\frac{\phi}{f}\right)+8 \alpha M^4+1\right)^{2/3}}
\end{equation}
\begin{equation}
\label{analytical_exp_r}
r = \frac{4  \sin ^{\frac{4}{3}}\left(\frac{\phi}{f}\right)}{\sqrt[6]{3} \sqrt[3]{\alpha} f^{4/3} \sqrt[3]{M^4} \left(\cos \left(\frac{\phi}{f}\right)+1\right)^{5/3} \left(4 \alpha M^4 \cos \left(\frac{\phi}{f}\right)+4 \alpha M^4+\frac{1}{2}\right)^{2/3}}
\end{equation}

\subsection{Comparison with observation}
In Fig. (\ref{caseII_caseIII}) we illustrate some acceptable points for a large value of $\alpha$. Whereas in slow roll scenario (section \ref{slow_roll_Inflation}) we fixed e-folding number $N$ and deduced the initial field value $\phi_*$ at horizon crossing where the spectral observables need to be evaluated, we shall proceed for the k-inflation scenario by giving $\phi_*$, then computing $N$ and verifying it is acceptable. Actually, there is an upper bound on $N$($\sim 64$) \cite{Linde0305263},  but it is, however, sensitive both to a possible reduction in energy scale during the late stages of inflation and to the complete cosmological evolution, so for some non-standard scenarios, one is permitted a higher N reaching, say, $80$ \cite{Linde0305263}.

The small (large) red circles indicate negative (positive) values of $\xi$. We took $\phi_*=65, f=45$ for all cases. For the negative $\xi$, we took ($\alpha=200, M^4=10^{-3}$) and we scanned $\xi$ in the interval $[-0.15,-0.05]$, and got $N$ acceptable, even though slightly high ranging from $72$, for the leftmost point just outside the acceptable spectral region, to $81$, for the rightmost point inside the acceptable spectral region. For the positive $\xi$, we took ($\alpha=250, M^4=2\times 10^{-3}$)  and we scanned $\xi$ in the interval $[+ 9 \times 10^{-6},15 \times 10^{-6}]$, and got  $N$ in the range $[71, 79]$

As said before, although such a scenario with `high' values of $N$ may not be plausible, but the analysis we presented just proved the viability of the model, albeit with a `relatively high' $N$.

We noted that larger values of $\alpha$ produce slightly smaller values of $r$, but that as long as $\alpha > O(10^3)$, the results remain largely identical. Also, we noted that  the model shows sensitivity to the values of $f$, in that for smaller values of $f$ the lines are displaced towards smaller values of $n_s$.

\section{Constant roll inflation: }
\label{constant-roll-inflation}
\subsection{Analysis}

In this section we discuss the model within the scenario of constant roll inflation in the limit $\alpha \ll 1$, where, unlike the slow roll case, we rather go one order further and work up to first order in $\alpha$. As a result, the action of Eq. (\ref{Action_w_XX}), after doing the change of variable of Eq. (\ref{phivar}), will take the form:
\begin{equation}
\label{Action_const_roll}
S=\int d^4x \sqrt{-g}\Big\{\frac{1}{2}R+ \mathcal{G}(Y,W) \Big\}
\end{equation}
where $\mathcal{G}(Y,W)=2 \alpha Y^2 - Y - W(\varphi)$ and $Y=\frac{1}{2}\partial_\mu \varphi \partial^\mu \varphi$.

The field equations corresponding to the above action are \cite{Odintsov:2019ahz}:
\begin{equation}
\label{Fried_1_constant}
6 H^2=\Dot{\varphi}^2+3\alpha \Dot{\varphi}^4+2 W(\varphi)
\end{equation}
\begin{equation}
\label{Fried_2_constant}
\Dot{H}=-\frac{\Dot{\varphi}^2}{2} -\alpha \Dot{\varphi}^4
\end{equation}
\begin{equation}
\label{EOM_varphi_const_roll}
12H \alpha (-\beta -1)\Dot{\varphi}^3 +2H (-\beta -3) \Dot{\varphi} - 2 W_\varphi =0
\end{equation}
where $\beta = \frac{\ddot{\varphi}}{H \dot{\varphi}}$ taken to be constant during the inflation.

Solving \eqref{EOM_varphi_const_roll} algebraically, one can find a real solution of $\Dot{\varphi}$ as,
\begin{equation}
\label{sol_dot_phi}
\Dot{\varphi}=\Big\{q(\varphi)+\big[q(\varphi)^2+p(\varphi)^3\big]^{1/2}\Big\}^{1/3}+\Big\{ q(\varphi)-\Big[q(\varphi)^2+p(\varphi)^3\Big]^{1/2}\Big\}^{1/3}
\end{equation}
where:
\begin{equation}
\label{coeff_dotphi}
q=+\frac{W_\varphi}{4\alpha H (1+\beta)} ,  p=\frac{3+\beta}{18 \alpha (1+\beta)}
\end{equation}
The slow roll parameters are given by:
\begin{equation}
\label{epsilon_const_roll}
\varepsilon=-\frac{3\Dot{\varphi}^2}{4W(\varphi)}(1+\alpha \Dot{\varphi}^2)
\end{equation}
\begin{equation}
\label{eta_const_roll}
\beta=\frac{\ddot{\varphi}}{H \Dot{\varphi}}
\end{equation}
\begin{equation}
\label{gamma_const_roll}
\gamma=\frac{6 \alpha \beta\Dot{\varphi}^2}{1+6 \alpha \Dot{\varphi}^2}
\end{equation}

For the slow-roll inflation, these slow parameters should satisfy the slow-roll conditions $( < 1 )$. However, in the constant-roll inflationary scenario, such conditions are not
necessary to be satisfied, as we only need $ \varepsilon < 1$ to achieve a significative inflation. The constant rate of roll is defined by imposing the constancy of the parameter $\beta$ defined eq. (\ref{eta_const_roll}) \cite{Motohashi:2014ppa}.

The spectral index and the tensor to scalar ration are given as \cite{Odintsov:2019ahz},
\begin{equation}
\label{ns_const_roll}
n_s=1+2\frac{2\varepsilon-\beta-\gamma}{1+\varepsilon}
\end{equation}
\begin{equation}
\label{r_const_roll}
r=E C_A^{3+\beta} \bigg(\frac{(1+6 \alpha \Dot{\varphi}^2)\Dot{\varphi}^2}{W(\varphi)}\bigg)
\end{equation}
where:
\begin{equation}
\label{coeffecient_const_roll}
E=\bigg(\frac{\sqrt{6}\Gamma(\frac{3}{2})}{\Gamma(\frac{3}{2}+\beta)2^\beta}\bigg)^2 ,
C_A=\frac{1-2\Dot{\varphi}^2}{1-6\Dot{\varphi}^2}
\end{equation}

In order to evaluate these spectral indices, and considering that expressing $\phi$ in terms of $\varphi$ via Eq. \ref{phivar} is not obtainable analytically, and so we are just equipped with the expression $W(\phi)$, we take as an input the filed $\phi$, rather than the canonical field $\varphi$. Using Eq. \ref{phivar}, so that to deduce $W_\varphi$ from $W_\phi$, and the approximation $H \sim \frac{W}{3}$, one is able to evaluate $\dot{\varphi}$ using Eqs. (\ref{sol_dot_phi} and \ref{coeff_dotphi}) which is necessary to get ($\varepsilon, \xi, n_s$ and $r$).

\subsection{Accommodation of observational data:}

In the limit $\alpha \ll 1$, the model appears as a kinematically derived inflation with the function $\mathcal{G}(Y,\varphi)=\alpha Y^2-Y-W(\varphi)$. However, unlike the limit $\alpha \gg 1$, where the studied model was also a k-inflation, the constant roll scenario is analyzed in the context of slow-roll k-inflation with the slow roll parameter $\beta$ remaining constant.  The input free parameters will be ($f, M, \alpha, \xi, \beta, \phi_* $), which will allow to compute ($n_s,r$ and $N$) and check whether or not they are acceptable.

\begin{figure}
\centering
\includegraphics[scale=.8]{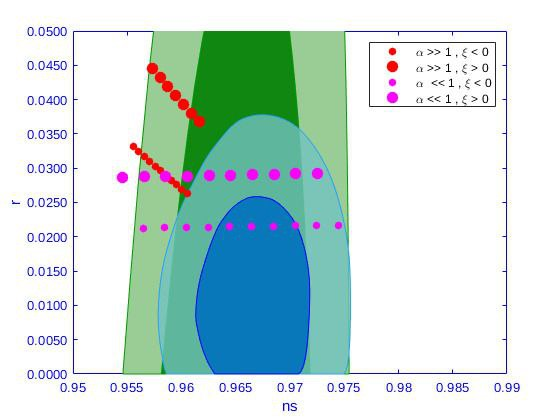}
\caption{\label{caseII_caseIII} Kinematically  derived  NI inflation, within F(R) gravity, via $\alpha R^2$-term, and NMC to gravity, through $\xi R \phi^2$-term.
For k-inflation with $\alpha \gg 1$ (red circles), we used ($\phi_*=65, f=45$). The small points correspond to ($\alpha=200, M^4=10^{-3}$) and to $\xi \in [-0.15, -0.05]$,  whereas we took ($\alpha=250, M^4=2\times 10^{-3}$) and $\xi \in [9 \times 10 {-6}, 15 \times 10^{-6}]$ for the large points. The obtained values of N correspond to spanning $[72, 81]$, going from the leftmost point to the rightmost one, for the small circles, and to spanning $[71, 79]$ for the large circles.
For constant-roll inflation, we took ($\alpha = 10^{-5}, f=9, M^4=11$) and scanned $\beta$ in the interval ($[0.010,0.019]$) for all pink points, with ($\xi=-6\times 10^{-4}, \phi_*=72$) for the small points and ($\xi=10^{-4}, \phi_*=78$) for the large ones. We got $N\sim 66.4$ ($N\sim 61.4$) for the pink small (large) points. The  models  are  contrasted  to  Planck \& BK  2018,  separately  or  combined  with  other experiments }
\end{figure}

The analysis is performed for different values of the new parameter $\beta$ as shown in Fig. (\ref{caseII_caseIII}). The small (large) pink points, which are acceptable regarding the spectral observables ($n_s,r$), correspond to negative (positive) values for the NMC constant. We took ($\alpha = 10^{-5}, f=9, M^4=11$) and scanned $\beta$ in the interval ($[0.010,0.019]$), where the lowest (largest) value corresponds to the leftmost (rightmost) pink dot, for all points, with ($\xi=-6\times 10^{-4}, \phi_*=72$) for the small points and ($\xi=10^{-4}, \phi_*=78$) for the large ones. We got a reasonable $N \in [66.4,66.5]$ and $N \in [61.3, 61.5]$ for the small and large points respectively.

We noted that  the model at this limit exhibited relatively high sensitivity to the inputs $\xi$ and $\beta$.

\section{Periodic NMC to gravity}
\begin{figure}
\centering
\includegraphics[scale=.5]{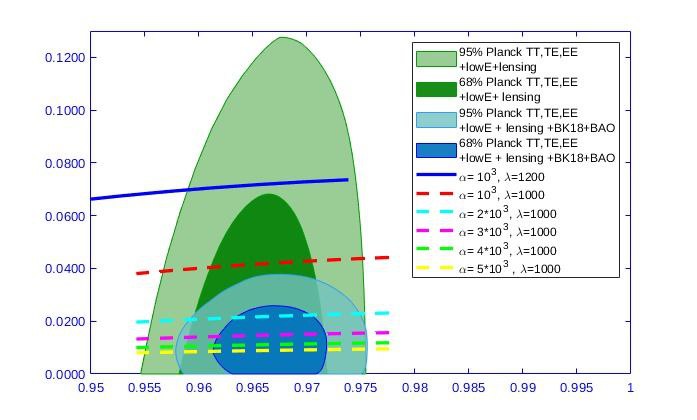}
\caption{\label{periodic_NMC} Contour plots of $n_s$ and $r$ for a periodic NMC to gravity of the form of Eq. \ref{periodic_NMC} (with $f= 140, V_0=10^3$) within an $\alpha R^2$-extended gravity. The NMC-parameter $\lambda$ takes two values $1000$ and $1200$, whereas we take several values for $\alpha$. The $\phi_*$ at horizon crossing scans the range $[265,280]$, and we find an e-folding $N$ in the range $[66,72]$.}
\end{figure}

In this section we consider a periodic NMC between the scalar field $\phi$ and the gravity of the form:
\begin{equation} \label{F_periodic_NMC}
F(\phi) = \lambda \left(\cos\left(\frac{\phi}{f}\right)+1\right)
\end{equation}
This type of coupling has been studied first in \cite{1806.05511}, where the authors assumed an NMC term $F(\phi)$  respecting the shift symmetry $(\phi \rightarrow \phi + 2 \pi f)$ and having a simple form proportional to the potential. The model was studied in the metric formalism with $\alpha=0$, and was shown to give rise to predictions for $n_s$ and $r$ that lay well within the $95\% $C.L. region from the combined Planck 2018+BAO+BK14 data. In \cite{2107.03389}, the author studied the periodic NMC within $\alpha R^2$ gravity, also within the metric formalism, and, in addition, presented a possible scenario where such a term emerges from a microscopic theory.
In the Palatini formalism, we find:
\begin{eqnarray}
 n_s &=& \frac{\mbox{Num}}{4 f^2 \cos ^2(\frac{{\phi}}{2 f}) \left({\lambda} \cos (\frac{{\phi}}{f})+{\lambda}+1\right)}: \\
 \mbox{Num}&=&  \left(f^2 (4 {\lambda}+2)+7 {\lambda}^2+2\right) \cos (\frac{{\phi}}{f})+{\lambda} \left(f^2+4 {\lambda}+1\right) \cos (\frac{2 {\phi}}{f}) \nonumber \\ && +3 f^2 {\lambda}+2 f^2+{\lambda}^2 \cos (\frac{3 {\phi}}{f})+4 {\lambda}^2-{\lambda}-6 \nonumber \\
  r &=& \frac{16 \tan ^2(\frac{{\phi}}{2 f}) \left({\lambda} \cos (\frac{{\phi}}{f})+{\lambda}-1\right)^2 \left({\lambda} \cos (\frac{{\phi}}{f})+{\lambda}+1\right)}{f^2 \left(4 \left(4 {\alpha} {V_0}+{\lambda}^2+{\lambda}\right) \cos (\frac{{\phi}}{f})+16 {\alpha} {V_0}+{\lambda}^2 \cos (\frac{2 {\phi}}{f})+3 {\lambda}^2+4 {\lambda}+2\right)} \\
 N &=& \frac{f^2 \left[2 \log \left(\tan (\frac{{\phi}}{2 f})\right)-\log \left(\tan ^2(\frac{{\phi}}{2 f})-2 {\lambda}+1\right)\right]}{2 {\lambda}-1}
\end{eqnarray}
We see in Fig. (\ref{periodic_NMC}) that the model is able to accommodate the observational constraints on $n_s$ and $r$, with a sufficient e-folding number. We find a similar effect to that of the quadratic monomial ($-\xi \phi^2 R$) in that when $\alpha$ increases then $r$ decreases with inconsiderate effect on $n_s$, whereas the increase of the coefficient in front of the field in the NMC term ($\lambda$ in Eq. \ref{F_periodic_NMC} and $-\xi$ in Eq. \ref{quadratic-monomial}) leads to a decrease in $n_s$, but now with a perceptible change on $r$ (look at the continuous blue line corresponding to $\lambda=1200, \alpha=10^3$ and the dashed red line with $\lambda=1000, \alpha=10^3$, where the former line is displaced to the left but higher than the latter). This again allows to put the ($n_s, r$) in acceptable regions, and we checked that the resulting e-folding number could be brought to acceptable values in $[65,72]$. As a representative point, for $(f=140, V=10^3, \lambda=1000, \alpha = 2000, \phi_*=265)$ we find $\phi_{\tiny \mbox{end}}= 437, N=69, r=0.0220$, and $n_s=0.9677$.

\section{Summary \& Conclusion}
In this work, we studied the NI with NMC between the inflaton field and gravity,  represented by the term $(-\xi \phi^2 R)$. We performed the work within an extended $R+\alpha R^2$ gravity under the Palatini formalism. Different cases of the model are examined within different scenarios of inflation. First, we investigated the model as a canonical slow roll inflation with re-scaled field and effective potential. We compared our results with others' in the literature and  showed that there are significant improvements regarding accommodation of data. Second, we examined the limit $\alpha \gg 1$, where the model is studied as K-inflation with Lagrangian of the form of Eq. (\ref{lagrangian_caseII_phi}). In a final stage, we considered the limit $\alpha \ll 1$ and the model was studied in the context of constant roll k-inflation with Lagrangian of the form of Eq. (\ref{Action_const_roll}). In  the second and third  cases,  we  showed  the  viability  of the  model  for  some  choices  of  the  free  parameters regarding  the  spectral  parameters$(n_s,r)$ and the e-folding number $N$. In each case, the theoretical consequences are compared to the results of Planck \& BK 2018 separately and  combined with other experiments. By studying the effects upon modifying the model, we found that the $\alpha R^2$-gravity influences the scalar-tensor ratio $r$ values. In contrast, NMC to gravity has a more significant impact on the spectral index values ($n_s$). Having contributions from both terms allows more previously excluded intervals, in the model free parameters space, to be included and to be compatible with observational data. Finally, we studied an NMC to gravity respecting the symmetry of the NI potential, so of the form ($\lambda(1+\cos(\phi/f))R$), and we saw that the conclusions of the quadratic monomial NMC hold somehow in this periodic NMC case, in that $\alpha$ affects, mainly, and in opposite directions $r$, whereas the NMC parameter ($-\xi, \lambda$) affect  $n_s$ in the same direction, but the change of $\lambda$ has a tangible effect on $r$ as well.

Whether or not this `focusing' ability to render the ($n_s,r$)'s points lie in a phenomenologically acceptable region, by playing the $\alpha R^2$ and the NMC's $\xi(\lambda)$-parameter, is a generic case, not specific just to NI, is a question that we hope this work will motivate further interest on its issue.

\section*{{\large \bf Acknowledgements}}
N.C. acknowledges support provided by the ICTP Senior Associate program, the CAS PIFI fellowship and the Humboldt Foundation.


\end{document}